\documentclass[onecolumn]{jpsj3}
\usepackage{txfonts}

\title{Commensurate and Incommensurate Vortex States Confined in Mesoscopic Triangles of Weak Pinning Superconducting Thin Films}

\author{Nobuhito Kokubo$^1$\thanks{E-mail: kokubo@pc.uec.ac.jp}, Hajime Miyahara$^1$, Satoru Okayasu$^2$, and Tsutomu Nojima$^3$}
\inst{$^1$Department of Engineering Science, University
of Electro-Communications, Tokyo 182-8585, Japan \\
$^2$Advanced Science Research Center, Japan Atomic Energy Agency, Ibaraki 319-1195, Japan\\
$^3$Institute for Materials Research, Tohoku University, Sendai
980-8577, Japan \\
} 

\abst{We report on the direct observation of vortex states confined
in equilateral and isosceles triangular dots of weak pinning
amorphous superconducting thin films with a scanning superconducting
quantum interference device microscope. The observed images
illustrate not only pieces of a triangular vortex lattice as
commensurate vortex states, but also incommensurate vortex states
including metastable ones. We comparatively analyze vortex
configurations found in different sample geometries and discuss the
symmetry and stability of commensurate and incommensurate vortex
configurations against deformations of the sample shape. }

\begin{document}
\maketitle


The concept of a small number of interacting particles in lateral
confinement is found in a variety of physical systems, including
electrons confined in semiconductor quantum dots
\cite{AshooriPRL1992}, electron dimples in the liquid-helium surface
\cite{LeidererSS1982}, and paramagnetic colloids in cavities
\cite{MangoldPCCP2004}. Of particular interest is vortex matter in
mesoscopic superconductors with different shapes
\cite{MoshNature1995, GeimNature1997, KandaPRL2003, CrenPRL2009}. In
addition to the vortex-vortex repulsion, vortices are subject to the
lateral confinement due to the shielding current flowing along the
sample boundary. The interplay between the intervortex interaction
and the confinement results in unique vortex states with strong
features of the sample shape, different from the
Abrikosov-triangular lattice in bulk superconductors
\cite{SchweigertPRL1998}. It is well established that in mesoscopic
disks, vortices form circular symmetric shells and obey the specific
rules for shell filling with increasing the vorticity $L$
\cite{BaelusPRB2004, MiskoPRB2007}. Meanwhile, in other geometric
shapes such as squares \cite{ZhaoPRB2008, MiskoSUST2009,
KokuboJPSJ2014} and pentagons \cite{HoSuST2013pentagon}, the
formation of vortex shells is not well defined as in disks owing to
the commensurability between the geometry and the vortex
arrangement. In mesoscopic triangles, which we focus on this study,
vortices form intrinsically a piece of the triangular lattice since
they match the geometric shape with threefold axial symmetry
\cite{ChibotaruPRL2001}. This occurs when vorticity becomes a
triangular number, $L = n(n+1)/2$ with integer $n$. For other
vorticities, the formation of the triangular arrangement is not
stable. Self-organized unique vortex patterns are formed
\cite{ZhaoEPL2008}. The corresponding vortex configurations are
numerically proposed to have twofold (middle-plane reflection)
symmetry and also to be represented by the combination of vortex
``bricks", i.e., lower vorticity configurations plus linear vortex
chains \cite{CabralPRB2009}.

The visualization of vortex states in triangular dots was initiated
by a scanning superconducting quantum-interference device (SQUID)
microscopy study \cite{kadowakiSTAM2005}, followed by Bitter
decoration studies \cite{ZhaoEPL2008, GrigorievaPRL2006}. Both
studies revealed some features dominated by inevitable inhomogeneity
(pinning sites for vortices) in samples (Nb films). To observe
vortex configurations inherent in mesoscopic triangles, it is
crucial to use nearly pin-free, weak pinning materials such as
amorphous superconducting films\cite{KokuboJPSJ2014, KokuboPRB2010}.
In this work, we report on the direct observation of vortices
confined in triangular dots of weak pinning amorphous MoGe films
with the scanning SQUID microscope.  Unlike the previous
experimental studies that were limited to dots with the regular
triangle shape \cite{ZhaoEPL2008, kadowakiSTAM2005}, we further use
isosceles triangular dots to obtain a better understanding of the
commensurability effect(s), and study the symmetry and stability of
commensurate and incommensurate vortex configurations against
deformations of the sample shape.

We used a commercial scanning SQUID microscope (SQM-2000, SII
Nanotechnology) with a dc SQUID magnetometer composed of Nb-based
Josephson junctions and an inductively coupled, pick-up Nb coil of
10 $\mu$m diameter integrated on a small Si chip
\cite{kadowakiSTAM2005}. The spatial resolution of the microscope is
$\sim$ 4 $\mu$m. The sample space is surrounded by a $\mu$-metal
shield, resulting in a residual magnetic field (ambient field) of $
\sim$ 1.2 $\mu$T. The details are described in Ref. 18.

\begin{figure}[t]
\includegraphics[width=17pc]{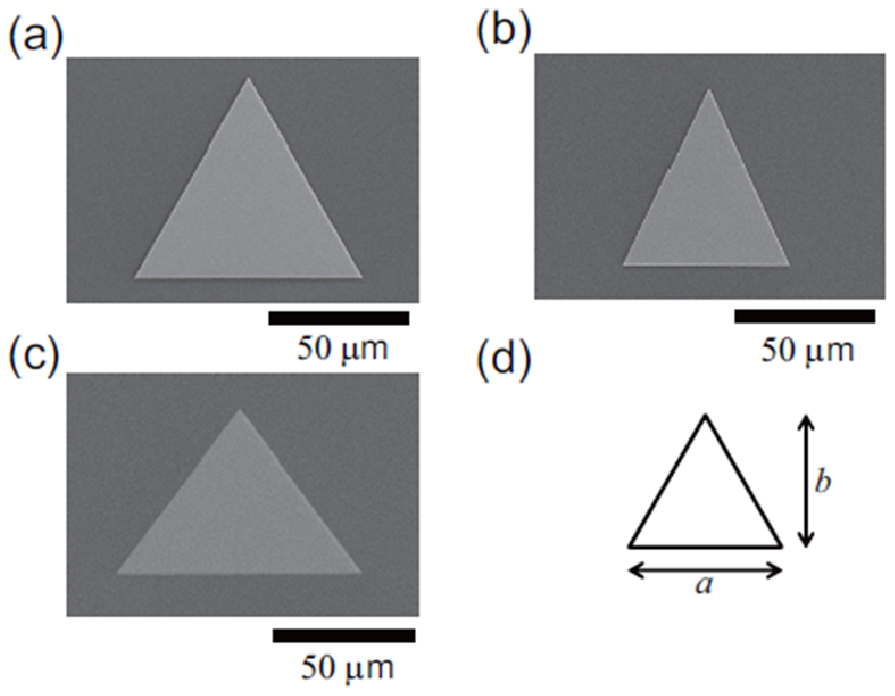}\hspace{2pc}
\caption{Fig. 1. Electron micrographs of triangular dots of
amorphous MoGe thin films. In addition to (a) equilateral triangular
dots, we made isosceles triangular dots of two different shapes. (b)
One (A) has a height expanded by 20$\%$ with respect to the regular
triangle, i.e., the ratio of the height $b$ with respect to its base
$a$ is $b/a=$1.04. (c) The other (B) has a height reduced by 20$\%$,
i.e., $b/a=$0.69.}
\end{figure}

Amorphous ($\alpha$-)Mo$_x$Ge$_{1-x}$ ($x\approx$ 78) films with
thickness $d=$0.20 $\mu$m were deposited on Si substrates by
rf-magnetron sputtering \cite{KokuboPRB2010}. The superconducting
transition temperature $T_c$ is $\approx$ 7 K, the coherence length
at zero temperature $T=0$ is $\approx$ 5 nm, and the magnetic
penetration depth $\lambda(0)$ at $T=$0 is $\approx$ 0.6 $\mu$m.
Since the films are thinner than $\lambda(0)$, the screening of the
supercurrent is characterized by the effective penetration depth
$[\Lambda=2\lambda^2/d \gg \lambda(0)]$ and the interaction between
vortices becomes long-ranged \cite{Nishio2008}. Using standard
photolithographic and wet-etching techniques, we fabricated
isosceles triangular dots of two different shapes, together with
equilateral triangular ones, as shown in Fig. 1. The isosceles
triangular A dots have a height expanded by 20$\%$ with respect to
the regular triangle, i.e., the ratio of the height $b$ with respect
to its base $a$ is $b/a=$1.04. The other B dots have a height
reduced by 20$\%$, i.e., $b/a=$0.69. We paid careful attention to
the selection of the dots to avoid vortex configurations induced by
pinning sites \cite{KokuboJPSJ2014}. We carried out a field-cooling
procedure for every SQUID measurement to obtain an equilibrium
vortex state \cite{KokuboPRB2010, FC}.

\begin{figure}[t]
\includegraphics[width=37pc]{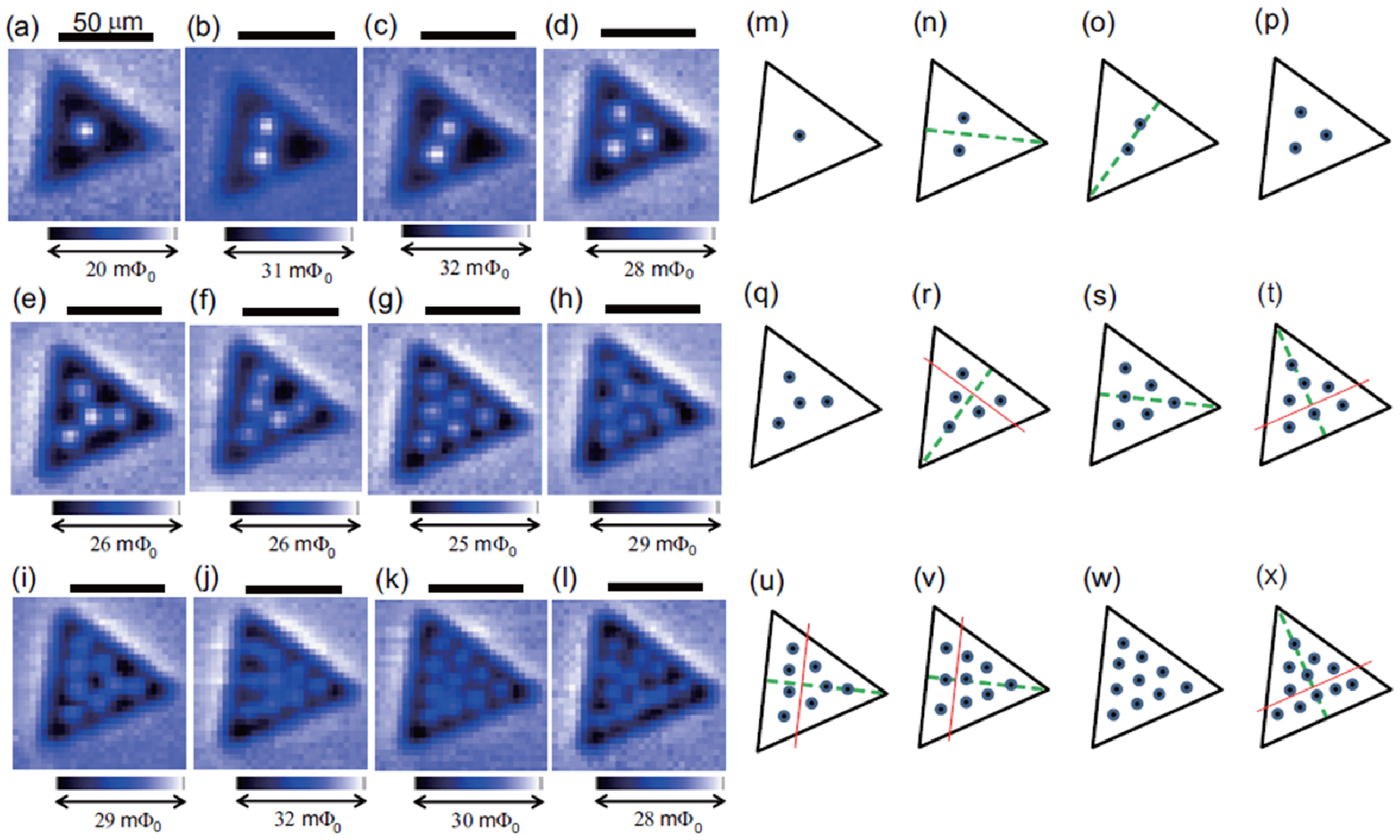}\hspace{2pc}
\caption{Fig. 2. Scanning SQUID microscopy images of vortices in a
75 $\mu$m equilateral triangular dot for vorticities $L =$ 1--11
observed after the dot was cooled to 3.7 K at different magnetic
(coil) fields of (a) 5.0, (b) 5.5, (c) 6.5, (d) 8.0, (e) 9.3, (f)
10, (g) 11.5, (h) 13, (i) 14, (j) 15, (k) 17, and (l) 18 $\mu$T. All
images are of areas of the same size (90 $\times$ 90 $\mu$m$^2$). A
color bar indicates the magnitude of the magnetic flux per pixel of
3$\times$3 $\mu$m$^2$ normalized by the magnetic flux quantum
$\Phi_0$. The traced vortex patterns are given in (m)--(x). The
broken lines represent symmetry axes of the triangular dot. The
solid lines are guides for the eyes.}
\end{figure}

Figures 2(a)--2(l) show vortex images in a 75 $\mu$m equilateral
triangular dot for vorticities $L =$ 1--11 observed after the dot
was cooled to 3.7 K under different magnetic fields. Each image
shows an area of 90$\times$90 $\mu$m$^2$ in size. A color bar
indicates the magnitude of the magnetic flux per pixel of 3$\times$3
$\mu$m$^2$ normalized by the magnetic flux quantum $\Phi_0$. The
corresponding traces of the vortex patterns are given in Figs.
2(m)--2(x). One can see the formation of pieces of the triangular
lattice at triangular numbers of vorticities $L=$ 3 [Fig. 2(d)], 6
[Fig. 2(g)], and 10 [Fig. 2(k)]. These configurations, together with
a $L=$ 1 configuration, are commensurate with the triangular dot
geometry and have the threefold rotational symmetry with respect to
the triangular dot center.  We also observe a commensurate
configuration at a nontriangular number of vorticity, $L=$ 4. As
shown in Figs. 2(e) and 2(q), a vortex appears in the dot center and
the other three vortices form a triangle. Namely, commensurate
patterns of $L=$ 1 and 3 are combined. We note that this is the only
case, and no other combination of commensurate patterns is observed
up to $L=$ 11.

For other nontriangular numbers of vorticities, we find
incommensurate vortex states with configurations that are frustrated
with the dot geometry. At $L=$ 2 [Figs. 2(b) and 2(c)], vortices
form a pair, but the pair orientation is rotatable in the dot.
Focusing on three axes of symmetry in the triangular dot, we find
that the observed orientations can be separated into two. One is
characterized by the orientation parallel to one of three symmetry
axes, as represented by a broken line in Fig. 2(o). The other is
perpendicular orientation, as shown in Fig. 2(n). As discussed
later, these two pair states are intrinsic to the triangular dot,
not induced by pinning sites.

At $L=$ 5 [Fig. 2(f)], vortices are aligned parallel to the edges of
the triangular dot and they form a V-shaped pattern. The observed
configuration is nearly symmetric with respect to one of three
symmetry axes in the triangular dot, as shown in the corresponding traced pattern [Fig. 2(r)]. 
It is interesting to note that this pattern can be divided into two:
one is a commensurate triangular pattern and the other is a linear
chain of two vortices. Other frustrated configurations at $L =$ 7--9
and 11 are also explainable in terms of the combination of a
commensurate vortex pattern and a linear vortex chain. The observed
configurations (except for the perpendicular orientation of a vortex
pair at $L=$ 2) are in excellent agreement with the lowest-energy
(ground) state configurations found in the numerical simulation
\cite{CabralPRB2009}.

Table I. Vortex configurations observed in equilateral and isosceles
triangular dots.
\begin{table}[htb]
  \begin{tabular}{cccc} \hline \hline
    $L$ &  \multicolumn{3}{c}{Configuration}  \\
    \cline{2-4}
      & Equilateral Triangle & Isosceles Triangle A  & Isosceles Triangle B \\\hline
    1 & $c1$ & $``c1"$ & $``c1"$ \\
    2 & $l2_{//}$, $l2_{\bot}$ & $l2_{//}$ & $l2_{\bot}$ \\
    3 & $c3$ & $``c3"$ & $``c3"$ \\
    4 & $c4$ & $``c4"$ & $``c4"$ \\
    5 & $c3+l2$ & $``c3"+l2$ & $l2+l3$ \\
    6 & $c6$ & $``c6"$ & $``c6"$ \\
    7 & $c4+l3$, $c3+l4$ & $``c4"+l3$ & $``c3"+l4$ \\
    8 & $c4+l4$ & $``c3"+l2+l3$ & $``c4"+l4$ \\
    9 & $c6+l3$ & $``c6"+l3$ & $l2+l3+l4$ \\
    10 & $c10$ & $``c10"$ & $``c10"$ \\
    11 & $c6+l5$ & $``c4"+l3+l4$ & $``c6"+l5$ \\ \hline \hline

  \end{tabular}
\end{table}

We summarize the observed vortex configurations for $L=$ 1--11 in
Table I using the notation for vortex configurations proposed in
Ref. 17. The $L=$ 5 state, for instance, can be denoted as $c3+l2$,
where $c3$ represents a commensurate $L=$ 3 pattern and $l2$ denotes
a linear chain of two vortices. Other incommensurate configurations
can also be characterized by this nomenclature. In most cases, each
$L$ is characterized by the ground-state configuration. However, at
$L=$ 7, we occasionally observe a \emph{metastable} $c3+l4$
configuration in addition to a $c4+l3$ configuration. Namely, the
ground and metastable states are observable (or nearly degenerated)
at $L=$ 7. The situation is similar to that for $L=$ 2 where two
pair orientations appear. For distinction, we add subscript ``$//$"
or ``$\bot$" for the parallel or perpendicular orientation,
respectively. The $l2_{//}$ configuration is numerically shown to be
the ground state \cite{ZhaoEPL2008, CabralPRB2009}, suggesting the
other ( $l2_{\bot}$) to be a metastable state.

We would like to emphasize that all the frustrated vortex
configurations observed in incommensurate states are characterized
roughly by the reflection symmetry with respect to one of three
symmetry axes in the triangular dot, in contrast to the threefold
rotational symmetry in commensurate states. The appearance of the
twofold symmetry in incommensurate states has also been shown
numerically \cite{CabralPRB2009}; however, the physical reason for
the symmetry is not obvious and requires further investigation. In
the following sections, we focus on the fact that the twofold
symmetry is the characteristic one of isosceles triangles, and study
whether the observed configurations appear in isosceles triangular
dots of two different shapes (isosceles A and B dots).

\begin{figure}[t]
\includegraphics[width=36pc]{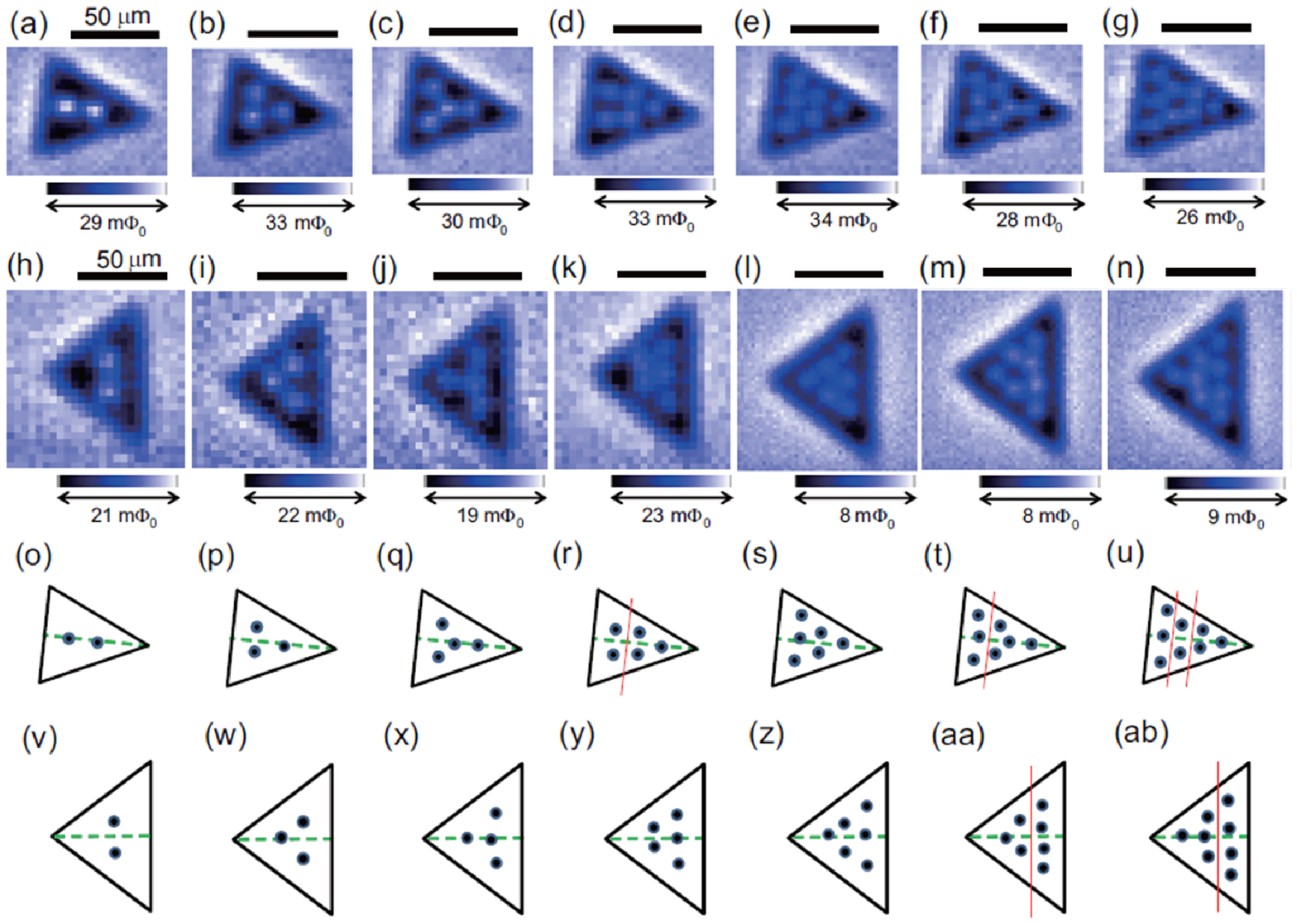}\hspace{2pc}
\caption{Fig. 3.  SQUID images of vortices in an isosceles
triangular A dot with 60 $\mu$m base for $L=$ 2--8 observed after
cooling to 3.6 K at different magnetic (coil) fields of (a) 8.0, (b)
10, (c) 12, (d) 14, (e) 15, (f) 16, and (g) 17 $\mu$T. These images
are of areas of the same size of 90 $\times$ 72 $\mu$m$^2$.  A color
bar indicates the magnitude of the magnetic flux per pixel size of
3$\times$3 $\mu$m$^2$ normalized by $\Phi_0$. The traced vortex
patterns corresponding to the images in (a)--(g) are given in
(o)--(u), respectively. We also present SQUID images observed in an
isosceles triangular B dot with 90 $\mu$m base after cooling to 2.7
K at different fields of (h) 2.0, (i) 2.5, (j) 3.0, (k) 3.8,(l) 4.5,
(m) 4.8, and (n) 5.0 $\mu$T.  These images are also of areas of the
same size of 100 $\times$ 100 $\mu$m$^2$. A color bar indicates the
magnitude of the magnetic flux per pixel size of 4$\times$4
$\mu$m$^2$ for (h)--(k), and that of 2$\times$2 $\mu$m$^2$ for
(l)--(n). The traced patterns corresponding to the images in
(h)--(n) are given in (v)--(ab), respectively. }
\end{figure}

Selected images observed in isosceles A and B dots are given in
Figs. 3(a)--3(g) and 3(h)--3(n), respectively. As traced in Figs.
3(o)--3(ab), all the vortex configurations are found to have the
twofold reflection symmetry with respect to the symmetry axis of
isosceles triangular dots. From the comparison of vortex images of
isosceles A and B dots, for each $L$, one can see how the vortex
configuration changes according to the geometry of the dots. At $L=$
3, for instance, three vortices in the isosceles A dot form a
triangle pattern, which is elongated along the symmetry axis of the
dot [Figs. 3(b) and 3(p)]. Meanwhile, a compressed triangle pattern
appears in the isosceles B dot [Figs. 3(i) and 3(w)]. If the
expansion and compression of triangle patterns are disregarded, the
vortex configuration is essentially unchanged. In other words, the
configuration is stable with respect to the deformation of the dot
shape. The situation is similar to other configurations observed at
$L$= 4 [Figs. 3(c), 3(j), 3(q), and 3(x)], 6 [Figs. 3(e), 3(l),
3(s), and 3(z)], and 10 (not shown). Thus, we identify those vortex
configurations (including $L=$ 1) as ``commensurate" states.

In other vorticities, vortices change their configurations according
to the geometry. At $L=$ 5, for instance, the vortex configuration
in the isosceles A dot seems to be viewed as the combination of a
$L=$ 3 triangular pattern and a linear chain of two vortices [Figs.
3(d) and 3(r)]. Meanwhile, in the isosceles B dot, it seems to be
constituted by two linear vortex chains [Fig. 3(k) and 3(y)]. The
differences in vortex configurations between isosceles A and B dots
are also noticeable at $L=$ 7 [Figs. 3(f), 3(m), 3(t), and 3(aa)], 8
[Figs. 3(g), 3(n), 3(u), and 3(ab)], 9 (not shown), and 11 (not
shown).

In Table I, we add vortex configurations observed in isosceles A and
B dots using the nomenclature discussed above. For distinction, we
append quotation marks in commensurate patterns in the isosceles
dots since they are determined from the stability with respect to
the geometric deformations, not by the (threefold axial)
configuration symmetry. One can find that the commensurate
vorticities in isosceles triangular dots (characterized only by
``commensurate" patterns) are $L=$ 1, 3, 4, 6, and 10, which
coincide with those found in equilateral dots. Thus, the stability
of vortex configurations against the geometric deformations leads to
the same conclusion with respect to the commensurability in vortex
states.

There are many coincidences in vortex configurations between
equilateral and isosceles dots. Of particular importance are
incommensurate states. At $L=$2, $l2_{//}$ and $l2_{\bot}$
configurations respectively appear in isosceles A and B dots, while
they appear as the ground and metastable states in equilateral dots.
The situation is similar to $c4+l3$ and $c3+l4$ configurations at
$L=$ 7. At $L=$ 5, equilateral and isosceles A dots share the same
configuration of $c3+l2$ as the ground state. More coincidences can
be found in other incommensurate vorticities ($L=$ 8, 9, and 11).
Thus, vortex configurations found in isosceles triangular dots are
intrinsically present in equilateral ones. These findings, together
with the configuration symmetry found in isosceles triangular dots,
provide strong support for the appearance of the twofold reflection
symmetry in the incommensurate vortex configurations in equilateral
triangular dots.

In summary, we have presented SQUID images of vortices in
equilateral and isosceles triangular dots of weak pinning
$\alpha$-MoGe thin films for $L$ up to 11. The observed images have
clearly shown the formation of pieces of a triangular lattice at
triangular numbers of vortices, $L=$ 3, 6, and 10. These
configurations, together with $L=$ 1 and 4 configurations, have the
threefold rotational symmetry and are essentially stable with
respect to the deformations of the dot geometry, i.e., the
transformation from equilateral triangles to two types of isosceles
ones by reducing or expanding the height with respect to its base.
Thus, we identified them as commensurate states. Meanwhile, for
incommensurate states, vortex configurations vary sensitively with
the geometric deformations. For each incommensurate $L$, the vortex
configurations observed in isosceles triangular dots with different
shapes differ from one another, and one of them (or sometimes both)
coincides with what is observed in equilateral dots. It turns out
that the twofold reflection symmetry characterizes the frustrated
vortex configurations observed in both equilateral and isosceles
dots. These findings provide not only a clear distinction between
commensurate and incommensurate states, but also strong support for
the appearance of the twofold reflection symmetry in frustrated
vortex configurations in mesoscopic triangular dots.


N. K. acknowledges discussions with S. Okuma. This work was
supported by JSPS KAKENHI Grant Numbers 23540416, 26287075, and
26600011, Nanotechnology Network Project of the Ministry of
Education, Culture, Sports, Science and Technology (MEXT), and the
Inter-university Cooperative Research Program of the Institute for
Materials Research, Tohoku University (Proposal No. 14K0004).


\begin{thebibliography}{9}
\bibitem{FC} The magnetic field was applied at some temperatures ($\sim$12 K)
above $T_c$, followed by cooling of the samples in the magnetic
field to temperatures ($\sim$ 2.7 -- 3.7 K) far below $T_c$ at which
scanning SQUID measurements were carried out.
\end{thebibliography}
\end{document}